\documentclass[12pt]{article}
\usepackage{graphicx}
\usepackage{natbib}
\usepackage{cite}
\usepackage{amssymb}
\usepackage{sidecap}
\usepackage{color}
\usepackage{times}
\usepackage{geometry}
\usepackage[utf8]{inputenc}
\usepackage{enumitem,amssymb}
\newlist{thematic}{itemize}{8}
\setlist[thematic]{label=$\square$}
\usepackage{pifont}

\usepackage{subcaption}
\sidecaptionvpos{figure}{h}
\usepackage{multicol}
\usepackage{hyperref}
\usepackage{titlesec}

\let\OLDthebibliography\thebibliography
\renewcommand\thebibliography[1]{
  \OLDthebibliography{#1}
  \setlength{\parskip}{0pt}
  \setlength{\itemsep}{0pt plus 0.3ex}
}

\titleformat{\section}
{\normalfont\large\bfseries}{\thesection}{1em}{}
\titleformat{\subsection}
{\normalfont\normalsize\bfseries}{\thesubsection}{1em}{}
\titleformat{\subsubsection}
{\normalfont\normalsize\bfseries}{\thesubsubsection}{1em}{}
\titleformat{\paragraph}[runin]
{\normalfont\normalsize\bfseries}{\theparagraph}{1em}{}
\titleformat{\subparagraph}[runin]
{\normalfont\normalsize\bfseries}{\thesubparagraph}{1em}{}

\titlespacing\section{0pt}{12pt plus 4pt minus 4pt}{4pt plus 2pt minus 2pt}
\titlespacing\subsection{0pt}{12pt plus 4pt minus 4pt}{4pt plus 2pt minus 2pt}
\titlespacing\subsubsection{0pt}{12pt plus 4pt minus 4pt}{4pt plus 2pt minus 2pt}
\titlespacing\paragraph{0pt}{12pt plus 4pt minus 4pt}{4pt plus 2pt minus 2pt}

\def\be{\begin{equation}} 
\def\ee{\end{equation}}

\def\HI{\hbox{H~$\scriptstyle\rm I\ $}}

\def\gsim{\lower.5ex\hbox{\gtsima}} 
\def\lsim{\lower.5ex\hbox{\ltsima}} \def\gtsima{$\; \buildrel > \over 
\sim \;$} \def\ltsima{$\; \buildrel < \over \sim \;$} \def\prosima{$\; 
\buildrel \propto \over \sim \;$} \def\gsim{\lower.5ex\hbox{\gtsima}} 
\def\lsim{\lower.5ex\hbox{\ltsima}} 
\def\simgt{\lower.5ex\hbox{\gtsima}} 
\def\simlt{\lower.5ex\hbox{\ltsima}} 
\def\simpr{\lower.5ex\hbox{\prosima}}   
  
 \def\gtsima{$\; \buildrel > \over \sim \;$} 
\def\ltsima{$\; \buildrel < \over \sim \;$} 
\def\gsim{\lower.5ex\hbox{\gtsima}} 
\def\lsim{\lower.5ex\hbox{\ltsima}} 
\def\simgt{\lower.5ex\hbox{\gtsima}} 
\def\simlt{\lower.5ex\hbox{\ltsima}} 
\def\simpr{\lower.5ex\hbox{\prosima}}

\def\E3{{\cal E}_{\rm g}^{III}}

\def\Msun{\rm M_\odot}

\def\log{\rm Log}

\def\avchi{$\langle \chi_\mathrm{HI} \rangle$}

\begin{document}
\begin{raggedright}
\huge
Astro2020 Science White Paper \linebreak

A proposal to exploit galaxy-21cm synergies to shed light on the Epoch of Reionization \linebreak
\normalsize
  
\noindent \textbf{Thematic Areas:} \hspace*{60pt} $\square$ Planetary Systems \hspace*{10pt} $\square$ Star and Planet Formation \hspace*{20pt}\linebreak
$\square$ Formation and Evolution of Compact Objects \hspace*{31pt} $\boxtimes$ Cosmology and Fundamental Physics \linebreak
  $\square$  Stars and Stellar Evolution \hspace*{1pt} $\square$ Resolved Stellar Populations and their Environments \hspace*{40pt} \linebreak
  $\boxtimes$    Galaxy Evolution   \hspace*{45pt} $\square$             Multi-Messenger Astronomy and Astrophysics \hspace*{65pt} \linebreak
  
\textbf{Principal Author:}

Name: Anne Hutter
 \linebreak						
Institution:  Kapteyn Astronomical Institute, University of Groningen
 \linebreak
Email: a.k.hutter@rug.nl
 \linebreak
Phone: +31 50 3634090
 \linebreak
 
\textbf{Co-authors:} (names and institutions)
  \linebreak
  Pratika Dayal (Kapteyn Astronomical Institute, University of Groningen),
  Sangeeta Malhotra (NASA GSFC),
  James Rhoads (NASA GSFC),
  Tirthankar Roy Choudhury (NCRA-TIFR Pune), 
  Benedetta Ciardi (Max Planck Institute for Astrophysics),
  Christopher J. Conselice (University of Nottingham), 
  Asantha Cooray (University of California, Irvine),
  Jean-Gabriel Cuby (Marseille Observatory),
  Kanan K. Datta (Presidency University Kolkata),
  Xiaohui Fan (University of Arizona),
  Steven Finkelstein (The University of Texas at Austin),
  Christopher Hirata (The Ohio State University),
  Ilian Iliev (University of Sussex),
  Rolf Jansen (Arizona State University),
  Koki Kakiichi (Department of Physics and Astronomy, University College London),
  Anton Koekemoer (Space Telescope Science Institute),
  Umberto Maio (Leibniz-Institute for Astrophysics Potsdam), 
  Suman Majumdar (Indian Institute of Technology Indore), 
  Garrelt Mellema (Department of Astronomy and Oskar Klein Centre, Stockholm University),
  Rajesh Mondal (University of Sussex),
  Casey Papovich (Texas A\&M University),
  Jason Rhodes (NASA JPL),
  Martin Sahlén (Department of Physics and Astronomy, Uppsala University),
  Anna Schauer (The University of Texas at Austin),
  Keitaro Takahashi (Faculty of Advanced Science and Technology, Kumamoto University),
  Graziano Ucci (Kapteyn Astronomical Institute, University of Groningen),
  Rogier Windhorst (Arizona State University),
  Erik Zackrisson (Department of Physics and Astronomy, Uppsala University)



\end{raggedright}

\pagebreak
\section {Introduction}

The emergence of the earliest galaxies, a few hundred million years after the Big Bang, led to a rapid transformation of the homogeneous and largely featureless infant Universe into an increasingly complex system. 
Star formation in these galaxies produced the first photons capable of ionizing the neutral hydrogen (\HI) atoms in the intergalactic medium (IGM) starting the Epoch of Reionization (EoR) that marks the last major phase-change of the Universe. The past decade has witnessed the emergence of a concordance picture in which reionization ended within the first billion years of the Universe \citep{fan2006, planck2016}. {\it Despite this enormous progress, the sources, history and topology of reionization (whether reionization proceeded inside-out from the densest to the rarest regions or vice versa) remain key outstanding questions in the field of physical cosmology} \citep{dayal2018}. Over the next decade a number of facilities, most notably the Square Kilometre Array (SKA), that aim to detect \HI in the EoR through its 21cm (spin-flip) transition, will be crucial in shedding light on the propagation of ionized regions. However, establishing the veracity of the 21cm signal and understanding the global sources and topology of reionization will require combining 21cm data with that from the underlying galaxy population. One such ideal data set is provided by a class of galaxies known as Lyman-$\alpha$ Emitters (LAEs), detected by means of their Lyman-$\alpha$ (Ly$\alpha$) emission line at 1216\AA\, in the galaxy rest-frame.  Lyman-$\alpha$ Emitters are particularly appropriate both to infer the global IGM ionization state and to study the reionization topology. This is due to (i) the precise redshifts yielded by the Ly$\alpha$ line {\it that significantly reduce smearing of the signal when correlating galaxy positions and 21cm data}; and (ii) the preferential location of LAEs in highly ionized regions that result in a {\it clear negative correlation} with the 21cm signal \citep[e.g.][]{wyithe2007, vrbanec2016, hutter2017, heneka2017, kubota2018}. 
Deep slitless spectroscopy with WFIRST, deep narrow-band imaging surveys with the Subaru Hyper Suprime-Cam and CTIO Dark Energy Camera, and wide-field spectroscopy with $>$25m-class telescopes, such as the European Extremely Large Telescope (E-ELT),  Giant Magellan Telescope (GMT), and Thirty-Meter Telescope (TMT), will together afford an unprecedented combination of sensitivity, volume, and redshift coverage for studying LAEs in the epoch of reionization.   Different facilities offer complementary strengths for this science.  The $>$25m telescopes offer the most sensitive line spectroscopy, but are strongest for $z<7$ followup of already-identified sources, where wide field optical spectrographs are most effective.    SKA and WFIRST offer continuous redshift coverage, allowing the techniques we discuss to be extended well beyond $z=7$.   While a number of works have focused on Subaru-SKA synergies \citep{hutter2018, kubota2018}, here we focus on optimising the synergy between SKA and WFIRST observations to allow a complementary and competitive approach to shed light on both the reionization history and topology. Throughout this paper we assume the standard $\Lambda$CDM cosmology with parameter values of $\Omega_\Lambda=0.73$, $\Omega_m=0.27$, $\Omega_b=0.047$, $H_0=100h=70$km~s$^{-1}$Mpc$^{-1}$ and $\sigma_8=0.82$.


\section{Theoretical model}

We use a state-of-the-art model that couples a cosmological smoothed particle hydrodynamic (SPH) simulations run using {\sc gadget-2}, an interstellar medium (ISM) dust model \citep{dayal2010} and the {\sc pcrash} radiative transfer code \citep{partl2011} to jointly track the evolution of reionization (and hence the 21cm emission) and the corresponding LAE population \citep{hutter2014}. This unique framework can be used, both, to explore the power of combining SKA and WFIRST observations and to delineate the best survey strategies to maximise such synergies. Focusing on $z \sim 6.6$, deep in the reionization era, the hydrodynamical {\sc gadget-2} \citep{springel2005} simulation has a box size of $80h^{-1}$~comoving Mpc (cMpc) and contains $2\times1024^3$ dark matter and, initially, the same number of gas particles; these correspond to a dark matter (baryonic) mass resolution of $3.6 \times 10^7 h^{-1}\Msun$ ($6.3 \times 10^6 h^{-1} \Msun$). The simulation includes physical descriptions for star formation, metal production and feedback and uses a Salpeter \citep{salpeter1955} initial mass function (IMF) between $0.1-100\Msun$. For each ``resolved" galaxy, corresponding to a halo mass $M_h>10^{9.2}\Msun$ the intrinsic spectrum is derived by summing over all the spectra of its star particles using the stellar population synthesis code {\sc starburst99} \citep{leitherer1999}. The observed Ly$\alpha$ luminosity is computed accounting for attenuation by both ISM dust and IGM \HI. Starting from a fully neutral IGM (i.e. with a global \HI fraction \avchi$=1$), we then run the radiative transfer code {\sc pcrash} on the $z \sim 6.6$ snapshot until it is fully reionized.  This model yields 21cm brightness temperatures for all intermediate ionization states (\avchi$\sim 1-10^{-4}$). For brevity, here we focus on results using an ionizing photon escape fraction value of $f_{esc}=0.05$ in reasonable agreement with observations \citep{finkelstein2019}.  To approximate potential WFIRST sample properties, galaxies with a Ly$\alpha$ equivalent width $EW_\alpha=L_\alpha^\mathrm{obs}/L_c^\mathrm{obs} \geq 20$\,\AA\ and a Ly$\alpha$ luminosity $L_\alpha \gsim 10^{42.5} {\rm erg\, s^{-1}}$ are identified as LAEs for different \avchi values and we derive the differential 21cm brightness temperature fields from the respective ionization field following \citet{iliev2012}.
\begin{eqnarray}
\delta T_b (\vec{x}) &=& T_0\ \langle \chi_\mathrm{HI} \rangle \ \left[1+\delta(\vec{x})\right] \ \left[1+\delta_\mathrm{HI}(\vec{x})\right] 
\label{eq_Tb}
\end{eqnarray}
Here, $1+\delta(\vec{x})=\rho(\vec{x})/\langle\rho\rangle$ and $1+\delta_\mathrm{HI}(\vec{x})=\chi_\mathrm{HI}(\vec{x})/\langle \chi_\mathrm{HI}\rangle$ refer to the local gas density and \HI fraction compared to their corresponding average global values, respectively. The fact that the 21cm-LAE correlations are derived for an un-evolving underlying galaxy population then allows us to disentangle the effects of reionization from galaxy evolution on LAE visibility. 


\section {Strategies to synergise WFIRST-SKA observations}

\begin{figure}[htb]
\center{\includegraphics[scale=0.63]{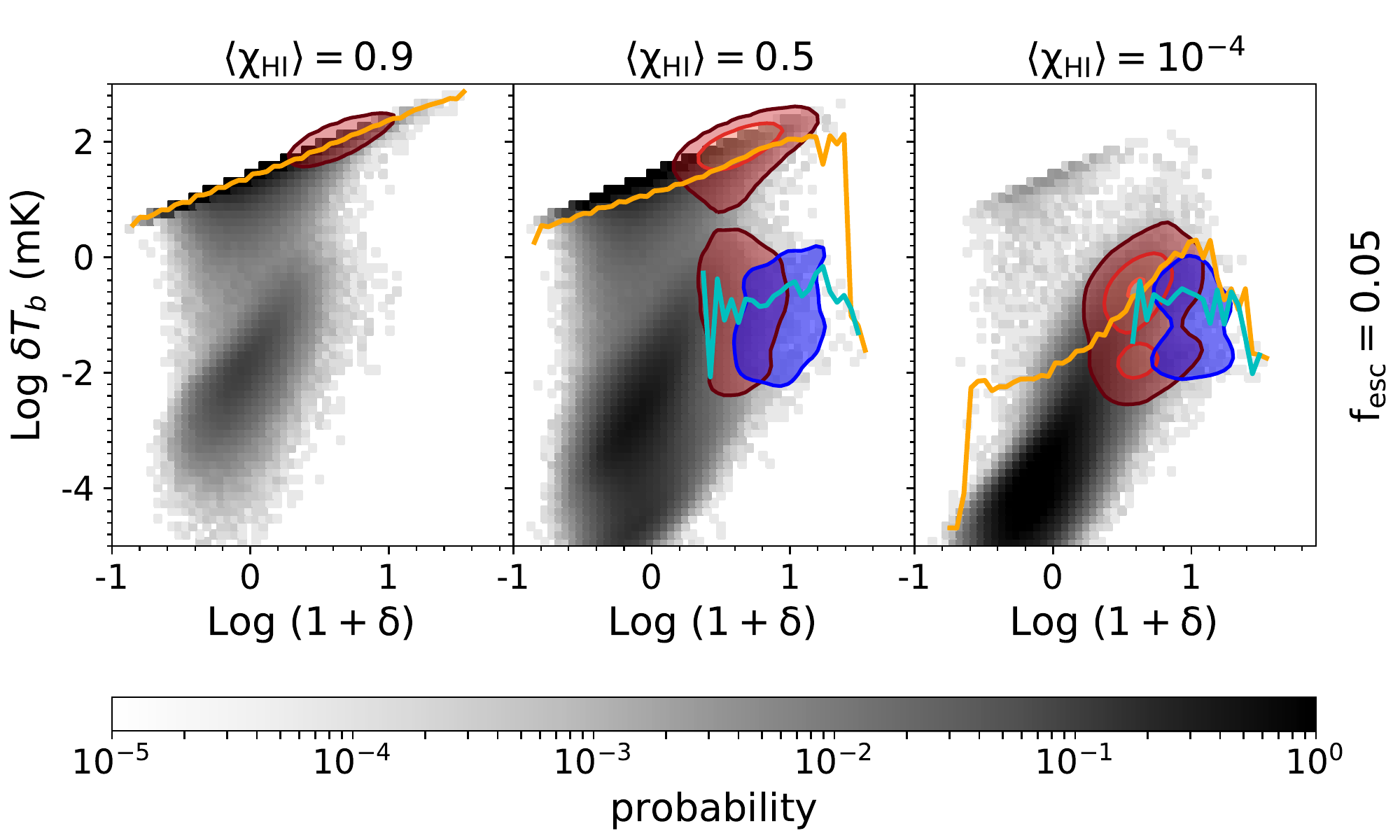}}
\caption{The probability density distribution of the IGM gas as a function of gas over-density ($1+\delta$) and the 21cm differential brightness temperature ($\delta T_b$) as the IGM progresses from being 90\% ({\it left panel}) to 50\% neutral ({\it central panel}) to fully reionized ({\it right panel}). The dark and light red contours show the regions occupied by 90\% and 50\% of all galaxies, respectively; the dark blue contours show the regions occupied by LAEs. The thick solid orange line shows the mean value of $\log(\delta T_b)$ for all cells; the cyan line shows the much lower mean $\log(\delta T_b)$ value in cells hosting LAEs.}
\label{fig1} 
\end{figure}

\begin{figure}
\center{\includegraphics[width=1.0\textwidth]{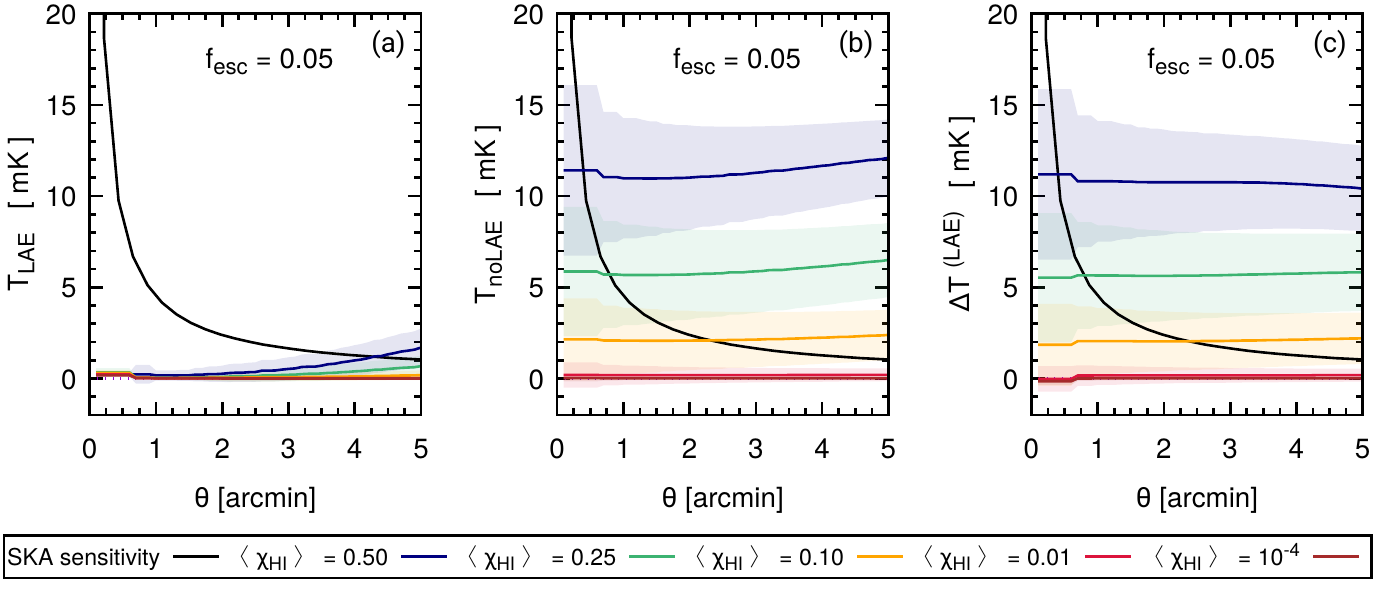}}
\caption{The differential 21cm brightness temperature in regions containing LAEs ({\it left panel}), in regions not containing LAEs ({\it central panel}), and their difference $\Delta T =  T_{\mathrm{noLAE}}-T_{\mathrm{LAE}}$ ({\it right panel}) as a function of the smoothing scale $\theta$. In each panel we show the differential brightness temperature at different stages of reionization (\avchi$=0.5$ - $10^{-4}$), as marked, and the black line shows the SKA imaging sensitivity limits for a 1000h observation. The difference between the brightness temperature in regions with and without LAES, easily testable using SKA, can unequivocally test the ``inside-out" topology of reionization predicted by models. }
\label{fig2} 
\end{figure}

\begin{SCfigure}
 \centering
  \includegraphics[width=0.48\textwidth]{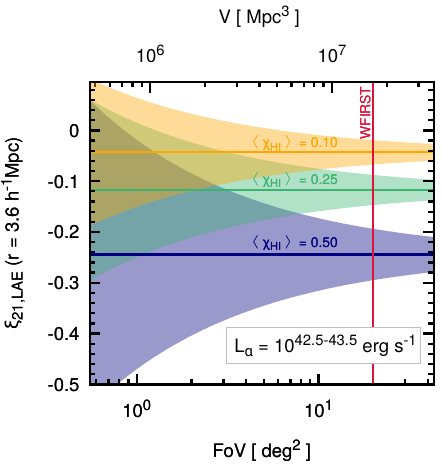}
  \caption{21cm-LAE cross correlation function at $r=3.6 h^{-1}$cMpc for a survey Ly$\alpha$ luminosity limit of $L_\alpha=10^{42.5}$erg~s$^{-1}$ for 1000h of SKA observations. The orange, green and blue lines represent results for \avchi~$\simeq0.1$, $0.25$ and $0.5$, respectively. The shaded regions show the cross correlation function uncertainties as a function of the survey volume of the SKA and LAE observations. The vertical line shows the survey area for WFIRST. Surveying an area of 20 deg$^2$ to a depth of  $L_\alpha=10^{42.5}$erg~s$^{-1}$, a correlation between WFIRST LAEs and SKA 21cm observations will be crucial in shedding light on the reionization state of the IGM.}
  \label{fig3} 
\end{SCfigure} 

\paragraph {The distribution of LAEs and 21cm emission:}

Given the large over-densities required for early galaxy formation, most galaxies lie in neutral regions in the initial stages of reionization (\avchi$\simeq 0.9$), as shown in Fig. 1.  By the time the IGM is half ionized, many galaxies (possibly those in clustered regions) are embedded in a fully ionised IGM with $\chi_{HI} \simeq 10^{-4}$, though some still occupy neutral regions ($\chi_{HI} \simeq 1$). Galaxies first become visible as LAEs when the IGM becomes (roughly) half ionised, since the Ly$\alpha$ flux gets attenuated by \HI in the IGM for higher \avchi~ values. We find $z \simeq 6.6$ LAEs to have halo masses $\gsim 10^{9.5}\Msun$ \citep{dayal2012}.   Observably bright LAEs at $z>6$ are among the high-mass end of the galaxy population, and lie in the most over-dense and highly ionized regions \citep{castellano2016}.  LAE hosting regions therefore have a much lower 21cm brightness temperature (compared to regions without LAEs) at any redshift where significant neutral gas remains. 

\paragraph {Hints on the topology of reionization:}

Given that observable LAEs occupy the largest halos in the most ionized regions, the brightness temperature in regions without LAEs ($T_{\mathrm{noLAE}}$) is generally higher than in regions hosting LAEs ($T_{\mathrm{nogal}}$), and both show a steady decrease as the IGM becomes increasingly ionized. In terms of SKA observations (Fig. 2), the brightness temperature difference between regions with and without LAEs ($\Delta T^{(LAE)}$) can be used to robustly differentiate between an IGM that is 10\% neutral to one that is 50\% ionized at these scales, irrespective of the $f_{esc}$ values used \citep{hutter2017}. The WFIRST-SKA synergy can be used to verify our prediction of an inside-out reionization scenario, where ionized regions percolate from over- to under-dense regions in the IGM, characterised by a lower differential 21cm brightness temperature in regions around LAEs compared to regions not containing LAEs. It also provides a unique tool to trace the typical size of ionized regions around LAEs by tracking the scale at which the 21cm-LAE cross power spectrum turns over \citep{lidz2009}.

\paragraph {Hints on the reionization state:}

To determine the best survey design for detecting the 21cm-LAE cross correlation $\xi_\mathrm{21,LAE}$ with SKA1-Low, we assume an SKA integration time of $1000$h and the array configuration V4A\footnote{\href{https://astronomers.skatelescope.org/wp-content/uploads/2015/11/SKA1-Low-Configuration_V4a.pdf}{\textit{https://astronomers.skatelescope.org/wp-content/uploads/2015/11/SKA1-Low-Configuration\_V4a.pdf}}}. We find that observational uncertainties will be critical in detecting the 21cm-LAE cross correlation signal and constraining \avchi. The key issue is that while uncertainties in the 21cm signal detection are reduced by larger survey volumes, the shot noise arising from a finite number of LAEs decreases as we probe to deeper Ly$\alpha$ luminosities \citep{hutter2018}. As shown in Fig. 3, LAE surveys with large fields of view ($\sim 20$ deg$^2$) and detecting LAEs with $L_\alpha \gsim 10^{42.5}\, {\rm erg \, s^{-1}}$ (corresponding to a flux limit of $6\times10^{-18}\, {\rm erg\, cm^{-2}\, s^{-1}}$ at $z\simeq6.6$), comparable to anticipated medium-deep WFIRST surveys carried out as parts of the WFIRST High Latitude Spectroscopic Survey, will be optimal to distinguish between an IGM that is $10$\%, $25$\% and $50$\% neutral.
Surveys that focus on LAEs brighter than at $L_\alpha\sim10^{43}$erg~s$^{-1}$ will be unable to exploit 21cm-LAE synergies, since LAE number densities become so low that the mitigation of the associated shot noise requires field of views exceeding that of SKA. It is important to note that, quantitatively, our results are quite robust to model assumption including the IMF, $f_{esc}$ and dust contents, given the degeneracy between these parameters \citep{dayal2011, hutter2014}. Finally, a combination of quasar data, from wide field surveys with, e.g., Euclid or WFIRST, and the 21cm emission detected by SKA will provide complimentary constraints on the local reionization state in the most over-dense regions at these early epochs \citep[e.g.][]{datta2016}.

\paragraph {Science Recommendation:} 

Over the next decade, 21cm experiments will aim to map out the progress of reionization. Synergising such observations with those of the underlying galaxy population (specially LAEs given their precise redshifts) will be absolutely crucial in shedding lights on the sources and propagation of reionization (i.e. if the process percolated from over- to under-dense regions or vice versa).  SKA and WFIRST offer a unique opportunity to apply these methods over an uninterrupted redshift range spanning the end of the reionization era.  We therefore recommend the utmost effort to maximize the synergy between SKA 21cm and WFIRST LAE observations, to finally understand the physics of the EoR that remains a crucial frontier in the field of astrophysics and physical cosmology. 

\paragraph{Ackowledgements:}
AH and PD acknowledge support from the European Research Council's starting grant ERC StG-717001 (``DELPHI"). PD also acknowledges support from the European Commission's and University of Groningen's CO-FUND Rosalind Franklin program. U.M. is supported through a research grant awarded by the German Research Fundation (DFG) project n. 390015701.

\renewcommand{\refname}{\normalfont\selectfont\normalsize\bf References:} 

\begin{multicols}{2}
\bibliography{mybib}
\bibliographystyle{mn2e}
\end{multicols}
\end{document}